\documentclass[prd,twocolumn,nofootinbib,preprintnumbers,amssymb,amsfonts,amsmath,superscriptaddress,showpacs,hyperref]{revtex4-1}

\usepackage{amssymb,amsmath,amscd}
\usepackage{graphicx,calc,epsfig,pstricks,bbm}
\usepackage{tikz}
\usepackage{enumerate}
\expandafter\ifx\csname package@font\endcsname\relax\else
 \expandafter\expandafter
 \expandafter\usepackage
 \expandafter\expandafter
 \expandafter{\csname package@font\endcsname}%
\fi

\newcommand {\be}{\begin{equation}}
\newcommand {\ee}{\end{equation}}
\newcommand{\ba}{\begin{array}{c}}
\newcommand{\ea}{\end{array}}

\begin{document}
\title{Cosmological singularity resolution from quantum gravity: \\ the emergent-bouncing universe}%

\author{Emanuele Alesci}
\email{emanuele.alesci@sissa.it}
\affiliation{SISSA, Via Bonomea 265, I-34136 Trieste, Italy, EU and INFN, Sez.~di Trieste.}

\author{Gioele Botta}
\email{gioele.botta01@universitadipavia.it}
\affiliation{Dipartimento di Fisica, Universit\`a degli Studi di Pavia, Via Bassi 6, I-27100 Pavia, Italy, EU}

\author{Francesco Cianfrani}
\email{francesco.cianfrani@ift.uni.wroc.pl}
\affiliation{Institute for Theoretical Physics, Uniwersytet Wroc\l{}awski, Pl.~Maksa Borna 9, Pl--50-204 Wroc\l{}aw, Poland, EU}

\author{Stefano Liberati}
\email{stefano.liberati@sissa.it}
\affiliation{SISSA, Via Bonomea 265, I-34136 Trieste, Italy, EU and INFN, Sez.~di Trieste.}

\date{\today}%

\begin{abstract} 
Alternative scenarios to the Big Bang singularity have been subject of intense research for several decades by now. Most popular in this sense have been frameworks were such singularity is replaced by a bounce around some minimal cosmological volume or by some early quantum phase. This latter scenario was devised a long time ago and referred as an ``emergent universe" (in the sense that our universe emerged from a constant volume quantum phase). 
We show here that within an improved framework of canonical quantum gravity (the so called Quantum Reduced Loop Gravity) the Friedmann equations for cosmology are modified in such a way to replace the big bang singularity with a short bounce {preceded by a metastable quantum phase in which the volume of the universe oscillates between a series of local maxima and minima}. We call this hybrid scenario an ``emergent-bouncing universe" since after a pure oscillating quantum phase the classical Friedmann spacetime emerges. Perspective developments and possible tests of this scenario are discussed in the end. 
\end{abstract}

\pacs{04.60.-m, 04.60.Pp, 98.80.Qc}

\maketitle

\section{Introduction}
Physical situations in which quantum gravity effects are expected to be relevant typically correspond to extreme gravitational regimes, such as the initial phase of the universe and the interior of black holes. As such, the characterisation of such regimes is crucial for the validation of quantum gravity (QG) models. 

The most promising scenario for experimental confirmation of quantum gravity phenomenology is realized in cosmology. The detection of cosmic microwave background radiation provides a unique window on phenomena occurring at early stages and, even though inflation washes out the initial conditions, some imprints of the quantum phase of the universe could survive in the radiation filling the universe today and be revealed in forthcoming experiments. 

There are two main scenarios replacing the Big Bang in QG: the bouncing universe and the emergent universe. The bouncing universe \cite{Novello:2008ra} predicts that going backward in time the universe collapses up to a finite value for the scale factor, before which it starts expanding again. The bounce implies a violation of the null energy condition and it has been realized only with some exotic matter fields, as in quantum cosmology \cite{Cai:2009zp}, or including modifications in the gravitational sector, as in Loop Quantum Cosmology (LQC)~\cite{Bojowald:2011zzb,Ashtekar:2011ni}. In this latter case, the granular spacetime structure makes gravity repulsive at Planckian scales, thus bridging the collapsing and the expanding classical solutions. 

The emergent universe scenario postulates that the universe started from a local static quantum phase. A realization has been proposed in String Gas cosmology \cite{Brandenberger:1988aj}, 
in the presence of a scalar field with an exponential potential \cite{Ellis:2003qz} and in Galileon cosmology \cite{Creminelli:2010ba}. Another realization of this scenario has been proposed in early developments of LQC \cite{Mulryne:2005ef}, using some quantum gravity corrections (the so-called inverse volume corrections), which later on were shown to be sub-dominant with respect to those producing the bounce (holonomy corrections).  

In an attempt to discriminate among such scenarios, we felt useful to investigate the outcome of some more refined scheme of calculation starting from quantum gravity.  In this sense the theory we consider is Quantum Reduced Loop Gravity (QRLG) \cite{Alesci:2013xd, Alesci:2015nja, Alesci:2016gub}, which provides an alternative realisation of the cosmological sector of Loop Quantum Gravity (LQG) with respect to LQC. 

While the latter implements a LQG-inspired quantisation scheme on a symmetry reduced classical kinematics (minisuperspace), in QRLG the symmetry reduction is achieved in two steps. Firstly, by implementing at the quantum level some convenient gauge fixing --- i.e.~the diagonality of the 3-metric (imposed {\em \`a la} Gupta--Bleuer) which greatly simplifies the quantum kinematics --- and then by imposing a semiclassical limit via the evaluation of the so obtained quantum Hamiltonian on coherent states implementing homogeneity and isotropy.

Let us stress that QRLG can be considered as an intermediate theory between full LQG and LQC, since while it preserves some of the basic LQG structures (e.g.~discrete graphs and intertwiners), it realises them in a simplified context in which computations can be performed analytically. In this way, it provides in the semiclassical limit, an Hamiltonian which can be seen as a quantum corrected form of the effective Hamiltonian of LQC.

In a previous work~\cite{Alesci:2016rmn}, two of the present authors demonstrated how the resulting universe evolution does not depart significantly from that of LQC when tracing back in time from the late universe up to the bounce.

Here, we study numerically the effective equations before the bounce, finding a surprising result: the pre-bouncing universe never becomes classical, but it emerges from a quantum phase.
{Such a quantum phase is characterized by oscillations between consecutive local minima and maxima of universe volume due to pure quantum gravity corrections. These oscillations, as we approach the classical expanding universe phase, tend to increase their amplitude, since as we go forward in time matter starts to dominate over gravity. Then after a last bounce similar to the one of LQC}, a super-inflationary phase occurs and eventually a classical Friedman Robertson Walker (FRW) universe develops. 

It is worth noting that this new scenario is not postulated at all, but it follows directly from solving some effective equations of motion derived via a non-perturbative treatment of the universe quantum gravitational degrees of freedom.  Hence our results provide a first realization of an emergent-bouncing universe scenario from a full fledged quantum gravity setting.

\section{Singularity resolution in LQC}

In order to understand the origin of the new physical scenario we are going to present here, it is convenient to sketch first how the resolution of the Big Bang singularity is realised in LQC. 

As briefly mentioned before, in this framework a LQG-inspired quantisation is applied on a phase space which has already been constrained to be homogenous and isotropic.  Let us then consider a flat FRW model (we adopt units $G=\hbar=c=1$) 
\be
ds^2=N^2(t)\,dt^2-a^2(t)\,(dx^2+dy^2+dz^2)\,,
\ee
$a$ and $N$ being the scale factor and the lapse function. It is convenient to parametrize the phase-space in terms of the following variables 
\be
b=\gamma\,\frac{\dot{a}}{a}\qquad v=\mathcal{V}_0\,\frac{a^3}{2\pi\gamma}  \, ,
\ee
$\gamma$ and $\mathcal{V}_0$ being respectively the Immirzi parameter and the considered fiducial volume\footnote{Note that such fiducial volume is just an infrared regulator, which can be safely removed in the end, {since all the results can be written in terms of physical variables $b$ and $v$}.}, while the only non-vanishing Poisson brackets reads
\be
\{v,b\}=-2\, .
\ee

The quantization procedure of LQG is inequivalent with respect to the Wheeler--DeWitt formulation, since one of the two phase space coordinates, being related to connection variables, cannot be promoted to a quantum operator.~\footnote{This is due to the fact that in LQG only holonomies operators exist and connection operators are not defined.}
As a consequence of this quantisation scheme, one finds that the effective classical Hamiltonian of LQC is~\cite{Ashtekar:2006wn,Singh:2012zc}
\be
H_{LQC}=-\frac{3v}{4\Delta\gamma}\,\sin^2(b\sqrt{\Delta})+\frac{P_{\phi}^2}{4\pi\gamma v}\,,
\label{ciao}
\ee
where $\Delta=4\pi\sqrt{3}$ equals the minimum eigenvalue of the area operator in LQG and $P_{\phi}$ denotes the momentum of a massless non-interacting scalar field used to model the universe thermal bath (this is the most relevant kind of non-exotic matter field close to the singularity). 

The Hamiltonian is constrained to vanish because of invariance under diffeomorphisms along the time-like direction. The basic difference with the classical Hamiltonian is due to the term $\sin^2(b\sqrt{\Delta})$, which replaces $\Delta b^2$. This replacement (holonomy correction) is a direct consequence of the adopted polymer-like quantization scheme \cite{Ashtekar:2002vh}. Inverse volume corrections are also present, but they are sub-dominant with respect to the holonomy ones and in what follows we will neglect them. 

Now the relevant  physics can be conveniently conveyed  in terms of effective equations of motion, since simulations have shown tha they are able to capture all the relevant quantum corrections~\cite{Ashtekar:2006wn,Singh:2012zc}. 
By computing the Hamilton equations and using the Hamiltonian constraint $H_{LQC}=0$, one gets 
\begin{align}
\left(\frac{\dot{a}}{a}\right)^2 &=\frac{8\pi}{3}\rho_{\rm m}\,\left(1-\frac{\rho_{\rm m}}{\rho_{\rm cr}}\right)\, ,\label{F1}\\
\left(\frac{\ddot{a}}{a}-\frac{\dot{a^2}}{a^2}\right) &=-8\pi\rho_{\rm m}\left(1-2\frac{\rho_{\rm m}}{\rho_{\rm cr}}\right),
\end{align}
where the energy density of the scalar field is $\rho_{\rm m}=P_\phi^2/2 V^2$ (with $V=\mathcal{V}_0 a^{3}$ equal to the physical volume) and $\rho_{\rm cr}\equiv{3}/{(8\pi\gamma^2\Delta)}$ is a critical energy density.

\begin{figure}
	\centering
	\includegraphics[width=0.9\linewidth]{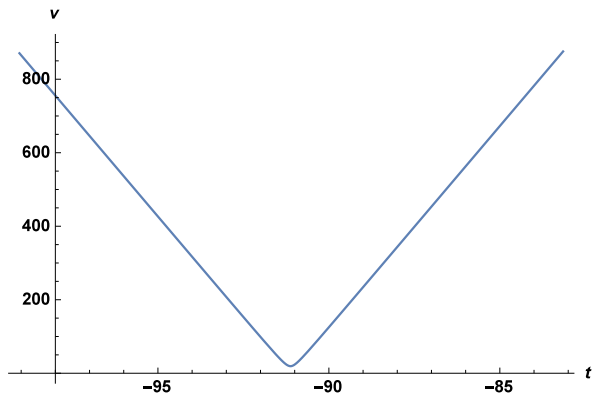}
	\caption{Dynamics of $v$ in LQC: the universe experiences a contracting phase after which undergoes to a bounce followed by an expanding phase.}
	\label{fig:lqc-}
\end{figure}

For $\rho_{\rm m}\ll \rho_{\rm cr}$ the evolution is essentially the one dictated by classical Friedmann equation, but for $\rho_{\rm m}\sim \rho_{cr}$ a bounce occurs and, going backward in time, the universe starts expanding again. The resulting picture is a bouncing universe, thus a resolution of the singularity (the equations of motion for $v$ and $b$ have been solved numerically and the result for $v$ is shown in Figure~\ref{fig:lqc-}).     

\section{Modified equations in QRLG}
Let us start by briefly review the procedure that leads to the effective Hamiltonian derived in \cite{Alesci:2016rmn}. This Hamiltonian is constructed as the expectation value of the quantum Hamiltonian proper of QRLG over  homogenous and isotropic semiclassical states and via a statistical average over an ensemble of classically equivalent systems: 
in QRLG states associated to the gravitational field are represented by cubical graphs. 

In this context semiclassical states have been constructed in the homogeneous and isotropic case by labeling different quantum states by a total number of cells $N$ and by peaking the variables of all cubical cells around the same quantum numbers ($j$,$\theta$). These quantum numbers determine the collective phase space variables ($v,b$). 

It is possible to verify that there exist a lot of different microstate ($N$,$j$,$\theta$) that realize the same macrostate ($v,b$). Assuming that all the possible microstates  are equiprobable we have to take into account the weight associated to each one on a fixed macrostate. This can be done by considering a density matrix whose coefficients take into account the occurence of the various microstates through a binomial distribution
\be
\rho_{v,b}=\sum_{N}\frac{1}{2^{N_{max}}}\left(\begin{matrix}N_{max} \\ N \end{matrix}\right)|N,j,\theta\rangle\langle N,j,\theta|.\label{dmat}
\ee
Considering the Gaussian approximation to the binomial distribution and taking the expectation value of the QRLG Hamiltonian over \eqref{dmat}, it follows that
\be
H\sim \int \frac{\sqrt{2}}{\sqrt{\pi N_{max}}}\,e^{-\frac{(N-N_{max}/2)^2}{N_{max}/2}}N^{2/3}\sin^{2}\left(\frac{v^{1/3}b}{N^{1/3}}\right)dN.\label{6}
\ee
Then through a saddle point expansion around the maximum of the weighting factor $N_{max}/2$ and identifying $(2/N_{max})^{1/3}=\sqrt{\Delta}/v^{1/3}$ we obtain the following Hamiltonian
\be
H_{QRLG}=-\frac{3v}{4\Delta\gamma}\,\sin^2(b\sqrt{\Delta})+\frac{P_{\phi}^2}{4\pi\gamma v} - \frac{ b^2\Delta^{3/2}}{24\pi\gamma^2}\cos(2b\sqrt{\Delta}),\label{ciao2}
\ee
which coincides with that of LQC, Eq.~\eqref{ciao}, up to the last term which is the next to the leading order term of the saddle point expansion. 


In \cite{Alesci:2016rmn} it was outlined how the dynamics generated by the Hamiltonian \eqref{ciao2} overlaps that of LQC up to the bounce, while here we want to extend that analysis further. Let us then introduce the following quantities 
\begin{align}
&\rho_{\rm g}=-\frac{\Delta^{3/2}b^2}{9 V}\, ,\label{SK}\\
&\bar{\rho}_{\rm cr}=-\frac{1}{\Delta}\, ,\\
&\Omega_{\rm g}=-\Delta\, \rho_{\rm g}=\frac{\rho_{\rm g}}{\bar{\rho}_{\rm cr}}\, ,\label{SE}\\
&\Omega_{\rm m}=\frac{\rho_{\rm m}}{\rho_{\rm cr}}\,.\label{V1}
\end{align}
The density $\rho_{\rm g}$ has no analogue in LQC and can be interpreted as a negative energy density source whose origin is purely quantum gravitational (as testified by the dependence on the minimal area $\Delta$). In what follows we call it geometrical energy density, keeping in mind that it is not really an energy density source, being proportional to $\dot{a}^2$. We also defined $\Omega_{\rm g}$ as the ratio of the geometrical energy density to what we call critical gravitational energy density $\bar{\rho}_{\rm cr}$, the reason will be clear in what follows. Similarly, we defined $\Omega_{\rm m}$ as the ratio of {the matter} energy density to the critical energy density of LQC. 

From $H_{QRLG}=0$, one finds
\be
\sin^2(b\sqrt{\Delta})=(\Omega_{\rm m}-\Omega_{\rm g})/(1-2\Omega_{\rm g})\,,\label{gio2}
\ee
and since the maximum and the minimum of the right-hand side are for $\Omega_{\rm m}+\Omega_{\rm g}=1$ and $\Omega_{\rm m}=\Omega_{\rm g}$, it follows  
that $0<\Omega_{\rm m}+\Omega_{\rm g}\leq 1$, from which 
$0<\Omega_{\rm m},\Omega_{\rm g}<1$.
These conditions tell us that $\rho_ {\rm cr}$ and $\bar{\rho}_ {\rm cr}$ are upper bounds for $\rho_{\rm}$ and $\rho_{\rm g}$, respectively. 

The equations of motion we get from \eqref{ciao2} read for $H_{QRLG}$
\begin{align}
&\dot{v}=\frac{3v}{2\sqrt{\Delta}\gamma}\sin(2b\sqrt{\Delta})\bigg(1+\frac{\Delta^2 b}{9\pi\gamma v}\cot(2b\sqrt{\Delta})-\frac{\Delta^{5/2}b^2}{9\pi\gamma v}\bigg)\label{vpunto}\\
&\dot{b}=-\frac{P_{\phi}^2}{\pi\gamma\ v^2}+\frac{b^2\Delta^{3/2}}{12\pi\gamma^2 v}\cos(2b\sqrt{\Delta})\,,\label{M1}
\end{align}
from which the following modified Friedmann equation can be derived
\begin{eqnarray}
\left(\frac{\dot{a}}{a}\right)^2=&\left(\frac{\textstyle 8\pi}{\textstyle 3}\rho_{\rm m}+\frac{\textstyle \rho_{\rm g}}{\textstyle \gamma^2}\right)(1-2\Omega_{\rm g})^{-1}\left(1-\frac{\textstyle \Omega_{\rm m}-\Omega_{\rm g}}{\textstyle 1-2\Omega_{\rm g}}\right)\nonumber\\&\left(1+\frac{\textstyle 2\Omega_{\rm g}}{\textstyle b\sqrt{\Delta}}\cot(2b\sqrt{\Delta})-2\Omega_{\rm g}\right)^2 .
\label{F4}
\end{eqnarray}

We can start analysing the above equation neglecting subdominant terms in a ${1}/{V}$ expansion. A posteriori, we shall show, numerically solving the equations \eqref{vpunto} and \eqref{M1}, that this simplification still provides a correct intuition on the general behavior of the scale factor. Hence, instead of \eqref{F4}, let us consider
\be
\left(\frac{\dot{a}}{a}\right)^2=\left(\frac{8\pi}{3}\\\rho_{\rm m}+\frac{\rho_{\rm g}}{\gamma^2}\right)(1-2\Omega_{\rm g})^{-1}\left(1-\frac{\Omega_{\rm m}-\Omega_{\rm g}}{1-2\Omega_{\rm g}}\right).\,\label{F5}
\ee

One can immediately notice that (differently from what happens in the case of LQC) even if $\rho_{\rm m}=0$ there is a nontrivial evolution of the scale factor, entirely due to quantum gravitational corrections. This seems very reasonable as one does not generic expect the matter field to dominate the universe evolution close to the would be classical big bang singularity.  
Let us then characterise the stationary points of Eq.~\eqref{F5} by considering the two conditions ensuring $\dot{a}=0$
\begin{itemize}
\item $\Omega_{\rm m}+\Omega_{\rm g}=1$ corresponding to $\sin^2(b\sqrt{\Delta})=1$ and qualitatively of the same kind of bouncing point found in LQC,
\item $\Omega_{\rm m}=\Omega_{\rm g}$ corresponding to $\sin^2(b\sqrt{\Delta})=0$ which is a novel possibility allowed for by the last term in Eq.~\eqref{ciao2}.
\end{itemize}
The presence of the second type of stationary point with respect to LQC suggests that the pre-bounce dynamics could be different (if these other stationary points are not dynamically excluded). 

The other equation we get from \eqref{vpunto} and \eqref{M1} reads
\begin{align}
\left(\frac{\ddot{a}}{a}-\frac{\dot{a}^2}{a^2}\right)=
&-\left(\frac{3}{\Delta\gamma^2}\sin^2(b\sqrt{\Delta})+4\pi\rho_{\rm m}\right)\cdot
\nonumber\\&\left(1-2\sin^2(b\sqrt{\Delta})\right)\,,\label{R3}
\end{align}
from which using \eqref{gio2} it follows that $\Omega_{\rm m}+\Omega_{\rm g}=1$ corresponds to a local minimum of the universe volume, while $\Omega_{\rm m}=\Omega_{\rm g}$ is a local maximum. 
 
\section{Numerical analysis}
{We now present the solution of the full equations \eqref{vpunto} and \eqref{M1}, obtained through a numerical analysis performed with Mathlab and checking the validity of the solution through the Hamiltonian constraint $H_{QRLG}=0$. 

The numerical investigation has been carried out taking as initial condition a classical late FRW universe and then evolving such universe backward in time. The result is plotted in Figure~\ref{fig:emerg}. The initial data of the presented simulation are $v=10^4$, $b=0.005$ and $P_{\phi}=154$, no significant qualitative modification occurs taking different values corresponding to another classic universe.

\begin{figure}
	\centering
	\includegraphics[width=0.9\linewidth]{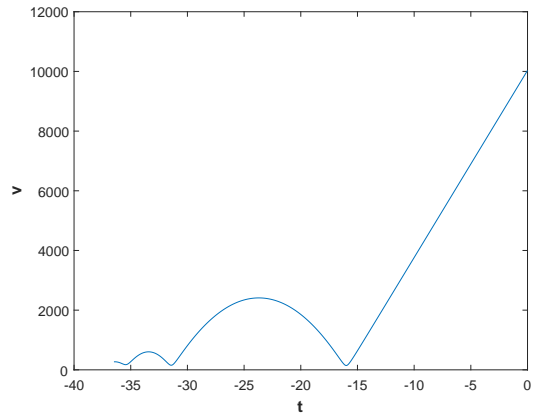}
	\caption{Dynamics of v in QRLG: Our classical universe emerges from a metastable quantum oscillating phase in which the volume of the universe is bounded and can never reach classical values}
	\label{fig:emerg}
\end{figure}

A careful reader might have noticed that in Figure~\ref{fig:emerg} the plot of $v$ has been truncated in the far past. Indeed, going backward in time the numerical simulation encounters a violation of the Hamiltonian constraint, and the saddle point expansion which leads to \eqref{ciao2} cannot be considered reliable any more. A more complete analysis of the full Hamiltonian constraints can indeed be carried out, and preliminary investigations seem to confirm that the scenario here presented can be extended without any qualitative change at arbitrary earlier times \cite{Paw}.
}

The behavior of the universe volume after the bounce confirms the results presented in \cite{Alesci:2016rmn}: it is qualitatively the same as in LQC up to the bounce, which corresponds to the condition $\Omega_{\rm m}+\Omega_{\rm g}\approx1$ and to a global minimum. 

However, the most interesting phase is the pre-bouncing phase shown in Figure~\ref{fig:emerg}. {Before the bounce, going backward in time, the universe experiences a metastable quantum phase in which the gravitational energy density and the matter energy density interact in such a way that the volume of the universe oscillate between the maximum condition $\Omega_{\rm m}\approx\Omega_{\rm g}$ and the minimum condition $\Omega_{\rm m}+\Omega_{\rm g}\approx1$. 

This signifies that, evolving backward in time, the gravitational energy density, Eq.~\eqref{SK}, starts to be non negligible and strongly modifies the time symmetric LQC bounce scenario. Indeed, being able to compensate the matter contribution, $\rho_g$ allows the volume of the universe in the prebounce phase to be bounded. 

Another interesting feature that emerges from Figure~\ref{fig:emerg} is the pre-bounce oscillatory behaviour: going forward in time the volume oscillations increase their amplitude, until the gravitational energy density is not able to compensate anymore the matter contribution and eventually the universe starts to expand approaching the classical Friedmann dynamics.}

{In summary, the final picture is that of a modified emergent universe scenario, where from an initial quantum phase the classical spacetime emerges, and the LQC bounce point, can be seen as the last minimum after which matter starts to dominates over gravity.} We can name this type of evolution an emergent-bouncing universe scenario.

%
%
\section{Discussion}
In this section, we discuss the physical origin of the prosed scenario. The novel dynamical feature with respect to LQC is provided by the last term in~\eqref{ciao2}. This term emerges as soon as the graph structure on which quantum states are based (and in particular the total number of nodes) is allowed to vary, according to the prescription discussed in~\cite{Alesci:2016rmn}. 

The summation over different graphs is a natural tool in LQG  and in this work we deduced its implications in a cosmological setting. It is evident that the most relevant of such implications is the explicit break of the scale invariance of the classical theory. 

{Indeed considering that homogeneity and isotropy are global properties of FRW cosmological spacetimes, the set of  global variables $(v,b)$ were introduced to represent a collective homogeneous and isotropic geometry on a quantum level. To do this, a global quantum number which characterize the whole graph was needed namely, the total number of nodes $N$. Hence it is quite natural to expect that the fluctuations of this quantum number lead to some global corrections, which in QRLG scale as $v_{pl}/v$. 

These corrections are different in nature from the local ones that lead to the bounce, indeed they can be considered a direct consequence of the lattice description of cosmological spacetimes in QRLG. It is worth noticing that this kind of terms, even if they are of a different physical nature, are also present in LQC, however they can be considered sub-leading corrections.} 

The appearance of this kind of corrections is considered problematic by some authors~\cite{Bojowald:2008ik}, essentially because it implies that the dynamics depends on how big the considered spatial region is. This is not surprising in a Quantum Gravity perspective, like QRLG and LQG, in which the atoms of space-time have fixed physical volume and, their number depends on the size of the considered spatial region, so that the dynamics is affected too. In other words, we are just saying that from a QG perspective it is quite reasonable that two Universes, one made by ten and the other by hundred Planckian cells, experience different dynamical behaviours.

Finally, it is also worth mentioning that the break of scale invariance is not at odds with the assumption of local homogeneity: homogeneity is preserved, in the form that all points in space ``see" the same cubic lattice.\footnote{This can be formally realised by taking semiclassical states for the geometry having some homogeneous quantum numbers and based at equivalence classes of diffeomorphism related graphs, such that the nodes of the graph are indeed abstract and not physical points.}    

\section{Conclusions}
We provided a new scenario for the resolution of the initial cosmological singularity, which mixes the bouncing and emergent scenarios, in a sense that the classical spacetime emerges from a pure quantum phase. {The universe starts from an early oscillating phase, but as matter starts to dominate over gravity, it undergoes to a final bounce and a rapidly expanding phase which approaches a classical FRW at late times. Quantum gravity corrections, computed in the framework of QRLG, are responsible for this peculiar behaviour: before the bounce they tend to compensate the matter energy contribution, bounding the volume of the universe to different local maxima that do not allow it to reach a classical regime.}

Let us stress that our derivation can be seen as a technical refinement on previous results obtained within Loop Quantum Cosmology (which seemed instead to suggest a time symmetric bounce) and as such it strongly suggests that once properly taking into account pure quantum gravitational (matter content independent) effects, an emergent universe scenarios could be a quite robust prediction within canonical quantum gravity.

Remarkably, the ``emergent-bouncing universe" scenario here derived, provides a completely new framework to address the problems of Standard Cosmology. {Since during the oscillating phase approaching the maxima and minima for the volume, the Hubble radius is infinite and the physical wavelength of fluctuations is approximately constant.} This solves the horizon problem. 

In this respect, it would be interesting to reconsider the so called entropy paradox and the generation of scale invariant perturbations within this framework as this might provide a testable alternative to inflation (see e.g.~the relevant discussion in~\cite{Brandenberger:2011gk}).
Alternatively, it would be worth investigating the modifications to the CMB spectrum predicted in this scenario. For example, by introducing an inflaton field realising inflation after the end of the bouncing phase, similarly to what has been done in LQC \cite{Ashtekar:2015dja}. 

Finally, it is worth mentioning that it has been claimed the quantum noise recorded by detectors at late time in the universe could be affected by the evolution of the latter at early times so that in principle experiments of this kind could be used to distinguish between big bang and bouncing scenarios~\cite{Garay:2013dya}. It would then be interesting to investigate the outcome of this sort of {\em gedankenexperiment} in the case of the hybrid emergent universe scenario here proposed (albeit one could suspect that the very similar behaviour around the bounce predicted by LQG and QRLG could leave a degeneracy between these models).

All of the above analyses could tell us how to experimentally distinguish the proposed model with respect to other bouncing and emergent scenarios. 
We hope that the present work will stimulate such exciting developments and further investigations which will provide tests for assessing the phenomenological viability of this new scenario.

\acknowledgments
FC is supported by funds provided by the National Science Center under the agreement
DEC-2011/02/A/ST2/00294. 
EA and SL wish to acknowledge the John Templeton Foundation for the supporting grant \#51876.

\end{document}